\documentclass[english,aps,pra,reprint,noshowpacs,superscriptaddress]{revtex4-1}   
\usepackage[T1]{fontenc}	% should generally be included for better accented-word behavior
\usepackage[latin9]{inputenc}	% should generally be included for better accent behavior
\usepackage{geometry}		% for controlling page margins
\usepackage{graphicx}
\usepackage[above,below]{placeins}	% allows use of \FloatBar­rier command to force section barriers
\usepackage{times}
\usepackage{hyperref}  
\hypersetup{colorlinks=true,urlcolor=blue,citecolor=blue,linkcolor=blue}   
\usepackage{tabularx}

\begin{document}

\newcommand{\insertFigure}[4]{
\begin{figure}
  	\includegraphics[width=#3]{#1}
  	\vspace{-2 em}
    \caption[#2]{#4\label{#2}}
    \vspace{-1.5 em}
\end{figure}
}

\newcommand{\mysection}[1]{
\vspace{-1.2 em}
\section{#1}
\vspace{-1 em}
}

\newcommand{\mysubsection}[1]{
\vspace{-1.4 em}
\subsection{#1}
\vspace{-1 em}
}

%\title{Views about experimental physics in a large introductory laboratory course}
\title{Transforming a large introductory lab course: impacts on views about experimental physics}
\author{Benjamin Pollard}
\affiliation{Department of Physics, University of Colorado Boulder, Boulder, CO 80309, USA}
\affiliation{JILA, National Institute of Standards and Technology, Boulder, CO 80309, USA}
\author{H. J. Lewandowski}
\affiliation{Department of Physics, University of Colorado Boulder, Boulder, CO 80309, USA}
\affiliation{JILA, National Institute of Standards and Technology, Boulder, CO 80309, USA}

%\keywords{}

\begin{abstract}
Laboratory courses are key components of most undergraduate physics programs. 
%In addition to reinforcing physics concepts, 
Lab courses often aim to achieve the following learning outcomes: developing students' experimental skills, engaging students in authentic scientific practices, reinforcing concepts, and inspiring students' interest and engagement in physics. 
Some of these outcomes can be measured by the Colorado Learning Attitudes about Science Survey for Experimental Physics (E-CLASS), a research-based assessment that measures students' views about experimental physics. 
We used E-CLASS at the University of Colorado Boulder to measure learning outcomes during a course transformation process in which views about experimental physics were reflected in the learning goals.
We collected over 600 student responses per semester from the large introductory laboratory course, both before and after implementing the course transformation. 
%We report on changes in E-CLASS responses from both the traditional and the transformed course.
While we observed no statistically significant difference in overall post-instruction E-CLASS scores before and after the transformation, 
in the transformed course, student responses to three E-CLASS items that related to the goals of the transformation were more favorable than in the original course.

%scores on three particular items shifted significantly towards more favorable responses in the transformed course.
%Those three items had been identified as related to our course transformation goals.
%However, compared to the original course, student responses to three E-CLASS items related to the goals of the transformation were more favorable in the transformed course.
\end{abstract}

\maketitle

\mysection{Introduction}
Science and technology are rooted in data and observation, and the field of physics is no exception.
Regardless of their subfield or specialization, a primary way experienced physicists evaluate their theories and ideas is by comparing them to empirical evidence \cite{Koponen2007}.
Furthermore, such evidence is often the progenitrix of new theories and ideas themselves \cite{Koponen2007}.
Therefore, physics students must gain an understanding of experimental techniques, equipment, processes, and habits of mind in order to better align their practices with those of expert physicists.
That learning occurs primarily in lab courses \cite{Caballero2018}.
As such, there is growing national interest in studying, investing in, and improving lab courses \cite{national2012discipline,kozminski2014aapt}.

In particular, lab courses are one of the only opportunities in undergraduate physics curricula for students to develop their views about experimental physics \cite{Wilcox2018}. 
%beleifs, attitudes, and expectations about experimental physics \cite{Wilcox2018}.
We define ``views about experimental physics'' to include attitudes, beliefs, and expectations about the nature and importance of experimental physics, as well as the strategies, emotions, and habits of mind that are involved when doing physics experiments.
%Such views are central to the subject-specific identities of practicing physicists, and are thus essential for students to develop \cite{Irving2014}.
Instructors of lab courses often mention the importance of students developing their views about experimental physics \cite{Dounas-Frazer2017}.
Consequently, learning goals of lab courses often include outcomes related to students' views about experimentation \cite{Dounas-Frazer2016b}.

With the above ideas in mind, we have led a research-based transformation of the introductory lab course at the University of Colorado Boulder (CU).
We identified a set of learning goals with input from STEM faculty at CU as an early step of the transformation process.
In addition to goals regarding measurement uncertainty \cite{Pollard2017,Lewandowski2017} and scientific communication, more than one learning goal focused explicitly on views about experimental physics.
Prior to the course transformation, we collected data from the original course for several semesters along dimensions aligned with these learning goals, in order to establish a baseline of comparison for the transformation.
The same data were collected in Spring 2018 during the first semester of the transformed course.

Here, we focus on the following research question: To what extent did our course transformation affect students' views about experimental physics?
To answer that question, we use data from the Colorado Learning Attitudes about Science Survey for Experimental Physics (E-CLASS), a research-based assessment tool for undergraduate lab courses \cite{Wilcox2018,Wilcox2017,Hu2017,Zwickl2014a}.
We compare 
%pre- and post-instruction 
E-CLASS results from the original and transformed courses.
%For context, we first describe the aspects of the course transformation relevant to our findings.

%In the future, these results will serve as a subset of the research documenting the impact of our transformation efforts.
%Overall, we expect that this work will identify strategies and areas of focus for effective teaching in physics labs, and illustrate the unique learning outcomes that are possible in lab courses.

\mysection{Context and methods}
%This work centers around a course transformation of the introductory lab course at CU.
%We are concerned here with the attitudes and beliefs about experimental physics of the students taking that course.
In this section, we provide details about (A) the course context both before and after transformation, (B) the E-CLASS assessment tool, and (C) our student population and methods.

\mysubsection{Course context}
The course studied here is the first physics lab course that students take at CU, typically in their second semester of study.
It is a large-enrollment stand-alone lab course worth one credit (typical courses at CU are worth three credits), usually taken concurrently with a separate physics lecture course.
% on electromagnetism.
The lab course, both before and after our transformation, consists of a series of weekly two-hour lab activities involving mechanics, electricity and magnetism, optics, and other physics concepts, all at an introductory level.
There is no midterm or final exam for the course; grades are based primarily on work produced from the weekly activities, including a pre-lab exercise, and participation in occasional supplemental lecture sessions.

Prior to the course transformation, students rotated through a series of lab activities, and different students worked on different activities at the same time.
%Students worked in pairs, and sometimes had the option of working individually.
Students performed measurements together, but were required to work independently when analyzing data and writing reports in order to reduce cheating.
Many of those activities involved measuring a known physical parameter already familiar to students, for example, using a pendulum to measure the acceleration due to gravity at the earth's surface.
%Each student was expected to write and turn in a lab report for each activity, due two weeks after the corresponding lab session. 
%Each report was scored by a graduate student teaching assistant.
%Those scores together represented the majority of each student's grade.

Our course transformation process was initiated in Spring 2016.
Faculty members and staff selected from physics, engineering, and applied science departments were interviewed to develop consensus learning goals for the course.
Five learning goals emerged, including ``Students' epistemology of experimental physics should align with the expert view,'' and ``Students should demonstrate an expert-like understanding of measurement uncertainty when evaluating measurements.''
Each learning goal was associated with an assessment tool, research-based if possible, for measuring corresponding learning outcomes.
%Beyond these stated learning goals, additional guiding principles emerged from the interviews, including ``Students will not write full formal lab reports,'' ``Students will not choose which experiments they do each week,'' and ``Students will be required to put in effort appropriate for a one-credit course,'' not more.
Further details of the course transformation process can be found in Ref. \cite{Lewandowksi2018}.

Each lab activity and lecture was completely redesigned as part of the course transformation.
The lectures were also updated to match the new lab activities and overall course content.
Nonetheless, many aspects of the transformed course remained the same or similar to the pre-transformed version, making the resulting course still ``traditional'' in the following ways.
Students still completed a pre-lab assignment before class, attended weekly lab sessions in which they worked through an activity guided by a manual, and attended occasional lectures.
%, and to include more examples from published professional scientific articles.

In the transformed lab sessions, all students did the same activity in a given week, mostly working in pairs or groups of three.
The activities were designed to encourage interaction between students, often prompting students to check with their partner or another group before continuing, and sometimes requiring students to combine their data with the rest of their lab section.
In most of the activities, students did not know the numerical result of their measurement before it was made. 
Instead, students were guided to make preliminary measurements of relevant parameters, and then to use those measurements to make predictions or measure unknown quantities.
For example, in one activity, students calibrated the change in capacitance of a parallel plate capacitor as known masses were placed on top of it, and then used those data to measure the weight of an unknown mass.
The activities often involved the explicit use of measurement uncertainty concepts, both to make predictions and compare data.
Instead of writing lab reports, 
%after their corresponding session, 
students created electronic lab notebooks during lab, and uploaded those notebooks before they left their session.

\mysubsection{E-CLASS assessment tool}
We use E-CLASS to measure student learning related to our transformation learning goals around attitudes and beliefs about experimental physics.
E-CLASS was developed in our research group at CU ``with the goal of providing a quantitative measure of the effectiveness of [course] transformations \cite{Wilcox2018}.''
The assessment consists of 30 statements about experimental physics, and asks students to rate their level of agreement with each statement on a 5-point Likert scale.
% from ``strongly agree'' to ``strongly disagree.''
Each statement has a defined expert-like response, confirmed and validated by the consensus of over 20 practicing experimental physicists and lab instructors \cite{Zwickl2014a}.
While the assessment asks students to respond both from their own perspective and from that of a hypothetical experimental physicist, we only consider the responses from the students' perspectives here.

E-CLASS measures students' beliefs, attitudes, and expectations about the nature of experimental physics.
This scope includes the importance of experimentation in the field of physics as a whole, the habits of mind employed while measuring and experimenting, students' emotional relationship with experimental physics, and the roles of common practices such as identifying research questions, making predictions, understanding equipment, fixing problems, interpreting raw data, and communicating results \cite{Wilcox2017}.
%The assessment started as an adaption of the Colorado Learning Attitudes about Science Survey (CLASS), and evolved through student and faculty interviews into its current, final, and validated form \cite{Zwickl2014a}.

\mysubsection{Student population and study methods}
The introductory physics lab course is one of the largest physics courses offered at CU.
Its enrollment has been steadily increasing over the past several years, with over 700 students in the course in Spring 2018.
Table \ref{demoTable} shows demographic information of the students enrolled in Spring 2018.
%In Spring 2018, 72.7\% of the students in the course identified as male, 25.6\% identified as female, and 1.7\% identified as non-gender conforming.
%That same semester, 67.1\% identified as white, 15.3\% as Asian American, 0.7\% as American Indian or Alaska Native, 2.3\% as Black or African American, 9.8\% as Hispanic/Latino, 0.4\% as Native Hawaiian or other Pacific Islander, and 4.3\% as an ``other race/ethnicity.''
%The vast majority of students in this course are not physics majors. 
%In Spring 2018, 17.9\% of students were physics majors, 43.0\% were engineering majors (excluding engineering physics), 35.1\% were non-physics STEM majors, and 4.0\% were majors in other disciplines.

\begin{table}[htbp]
\caption{Self-reported gender, race, ethnicity, and major of students enrolled in the course in Spring 2018.\label{demoTable}}
%\begin{ruledtabular}
\newcolumntype{R}{>{\raggedleft\arraybackslash}X}
\begin{tabularx}{\columnwidth}{R|r}
\hline \hline 
Female & 25.6\% \\
Male & 72.7\% \\
Gender non-conforming & 1.7\% \\
\hline
American Indian or Alaska Native & 0.7\% \\
Asian American & 15.3\% \\
Black or African American & 2.3\% \\
Hispanic/Latino & 9.8\% \\
Native Hawaiian or other Pacific Islander & 0.4\% \\
White & 67.1\% \\
Other race/ethnicity & 4.3\% \\
\hline
Physics & 17.9\% \\
Other Engineering & 43.0\% \\
Other STEM & 35.1\% \\
Other disciplines & 4.0\% \\
\hline \hline
\end{tabularx}
%\end{ruledtabular}
\end{table}
% Formatting tweak if needed--FloatBarrier forces floats to show here, before next section
%\FloatBarrier	

E-CLASS was administered online at the beginning and end of the semester, referred to respectively as pretest and post-test.
Data collection was performed for several years leading up to the first implementation of the transformed course.
Prior to the transformation, time was provided during regularly scheduled lab sessions for students to complete the assessment.
E-CLASS was also administered at the beginning and end of the transformed course in Spring 2018.
Due to time constraints, students in the transformed course completed the assessment outside of class using a link provided via email.
In all semesters, students received course credit for completing both the pretest and the post-test, amounting to about 1\% of their final grade in total.

In order to account for variations between Fall and Spring semesters, in this study, we use data from only the Spring 2017 semester (before the transformation) to compare to Spring 2018 (after the transformation).
In both of those semesters, the course was taught by the second author.

Data from E-CLASS was scored using established methods from previous studies \cite{Wilcox2017}.
The 5-point Likert scale responses to each E-CLASS item were collapsed into a 3-point scale, in which ``(dis)agree'' and ``strongly (dis)agree'' were combined into a single category. 
A response aligned with expert-like views was assigned a score of $+1$ and referred to as favorable, whereas a response opposite of expert consensus was assigned a score of $-1$ and referred to as unfavorable.
Neutral responses
%, ``neither agree nor disagree,'' 
were assigned a score of $0$.

Here, we analyze responses to individual E-CLASS items, and also compute an overall score for each student by summing the student's scores on all 30 E-CLASS items.
We compare the distributions of scores, both overall and item-by-item, using the nonparametric Mann-Whitney U test at the 5\% significance level to determine statistical significance.
To account for confirmation bias arising from performing multiple item-wise comparisons, we apply the Bonferroni correction, treating each item as an individual hypothesis.
We use Cohen's $d$ as a measure of effect size for statistically significant differences.
See Ref. \cite{Lomax2012} for a complete description.

Before analyzing any E-CLASS data from this course, both authors read through the list of 30 E-CLASS items and identified the ones that were related to our course transformation goals.
This subset of E-CLASS items was further discussed with the other instructor of the transformed course, resulting in a final set of nine items.
We consider only these nine items in our item-by-item analysis below.

\mysection{Results and discussion}
We begin by comparing the distributions of overall scores before and after the course transformation, comparing pretest to pretest and post-test to post-test.
The two distributions of overall pretest scores are statistically equivalent.
This result confirms that the student populations were similar in their views about experimentation at the start of the course.

Since the two pretests showed similar results, we turn to the post-tests to better understand how students' views changed by the end of the course.
We find that in overall score, the two post-test distributions are also statistically equivalent, suggesting that the course transformation did not have a marked effect on students' views about experimental physics when distilled into a single number.
However, the scope 
%of ``views about experimental physics'' 
that E-CLASS probes is far broader than our relatively narrow transformation learning goals. 
Thus, we do not expect our transformation to have a large effect on overall E-CLASS scores.
%\textit{This finding is also aligned with previous results from studies using CLASS and E-CLASS that showed student attitudes and beliefs tend to be overall stable in the course of a single academic semester \cite{stability_of_attitudes_and_beliefs}.}

%Similarly to above, we compare pretest to pretest and post-test to post-test, before and after the course transformation.
%However, differences in pretest item distributions have little practical significance, representing only uncontrolled differences arising from the distinct groups of students enrolled in Spring 2017 and Spring 2018.

\insertFigure{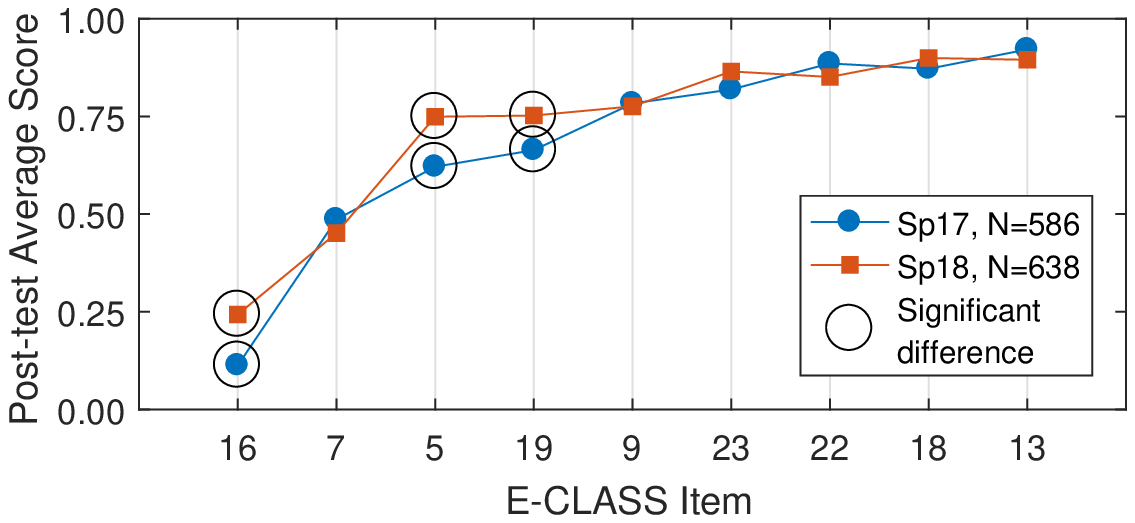}{postByQuestion}{\columnwidth}{
Comparison of post-test scores on nine E-CLASS items related to transformation goals, averaged across all students in the course.
Data from Spring 2017, before the transformation, is compared to data from Spring 2018, the transformed course.
Items are ordered by the mean of the scores from the two semesters.
Black circles mark the items that show statistically significant differences between semesters.
}

To get a more fine-grained perspective on the effects of the course transformation, we analyze E-CLASS post-test scores item-by-item.
The nine items related to our course transformation goals are shown in Table \ref{tabQuestions}, and plotted in Fig. \ref{postByQuestion}.
%Fig. \ref{postByQuestion} shows the average post-test score across all students in the course for each of the nine E-CLASS items that were identified as related to our course transformation goals.
Scores are plotted for both the Spring 2017 and Spring 2018 semesters.
%Note that the possible range of average scores on a single item is -1 to 1; all of these items were above 0, which corresponds to a favorable response.
Three of these items, those marked with black circles in the figure and bold in the table, showed statistically significant differences in post-test distributions before and after the transformation.
The $p$-values used to determine significance are shown for all items in the table.
The items with significantly different post-test distributions between the two semesters are highlighted in bold, and include corresponding $d$-values to measure effect size.

\begin{table*}[htbp]
\caption{E-CLASS items identified as related to our course transformation goals.
Items in bold show statistically significant changes before and after the transformation. 
Order of items is the same as in Fig. \ref{postByQuestion}.\label{tabQuestions}}
%\begin{ruledtabular}
\begin{tabularx}{\textwidth}{r|X|c|c}
\hline \hline 
Number & Statement & \textbf{$p$} & \textbf{$d$} \\ \hline 
16 & The primary purpose of doing physics experiments is to confirm previously known results. & \textbf{0.005} & 0.164 \\
7 & I don't enjoy doing physics experiments. & 0.339 &  \\
5 & Calculating uncertainties helps me understand my results better. & \textbf{0.001} & 0.207 \\
19 & Working in groups is an important part of doing physics experiments. & \textbf{0.005} & 0.156 \\
9 & When I approach a new piece of lab equipment, I feel confident I can learn how to use it well enough for my purposes. & 0.759 &  \\
23 & When I am doing an experiment, I try to make predictions to see if my results are reasonable. & 0.085 &  \\
22 & If I am communicating my results from an experiment, my main goal is to make conclusions based on my data using scientific reasoning. & 0.066 &  \\
18 & Communicating scientific results to peers is a valuable part of doing physics experiments. & 0.124 &  \\
13 & If I try hard enough, I can succeed at doing physics experiments. & 0.104 &  \\
\hline \hline
\end{tabularx}
%\end{ruledtabular}
\end{table*}
% Formatting tweak if needed--FloatBarrier forces floats to show here, before next section
%\FloatBarrier	

Among all nine items, we see no significant negative changes in average E-CLASS scores after the transformation, which suggests that the transformed course is not significantly worse than the original at promoting favorable views about experimentation along these nine dimensions.
Furthermore, five of the six items that remained unchanged after the transformation all have average scores above 0.75, indicating that these items 
%were favorable already and 
had little room to improve due to the course transformation.
The three items that did show significant change shifted towards more favorable responses after the transformation.
We discuss each in detail below.

Item 16 concerns confirming previously known results.
Note that the consensus expert response to this item is to disagree, as experiments usually serve a larger purpose than confirming previous results.
As a design goal aimed at achieving our learning goals, activities with the purpose of confirming previously known results were deliberately avoided in the transformed course.
The favorable shift in Item 16 after the transformation is likely related to this design goal.
Previous research has shown that students can interpret the phrase ``previously known results'' as either refering to their personal scientific knowledge, or to well-established results from the scientific community \cite{Hu2017}.
In either case, a shift towards more favorable responses on this item represents a desirable outcome of the course transformation.

Item 5 concerns the role of measurement uncertainty in interpreting results.
A more expert-like understanding of measurement uncertainty was a learning goal of the transformed course.
The statistically significant shift in Item 5 suggests that, distinct from the concepts themselves, students' views about measurement uncertainty were affected favorably by the transformation.
Previous work has shown that students in the original course already show pre-post shifts towards such conceptial understanding \cite{Pollard2017, Lewandowski2017}. 
%Further analysis is needed to know how students' views about measurement uncertainty and their conceptual understanding of it are related.
Additional work is needed to see whether conceptual understanding also improves in the transformed course and, if so, how students' views about and their understanding of measurement uncertainty are coupled.

Lastly, Item 19 concerns the importance of working in a group when doing physics experiments.
Group work was required in the transformed course, in stark contrast to the focus of the original course on individualized work.
%more strongly emphasized, in fact more often required, in the transformed course, which 
Furthermore, the activities in the transformed course often prompted students to compare and combine data with the entire class, promoting interactions between students specifically about their measurements.
These aspects plausibly contributed to the significant favorable shift in this E-CLASS item.

%Items 6 and 14 also showed significant shifts toward more favorable responses in the transformed course.
%While they did not correspond to the explicit learning goals or design principles of the team transforming the course, item 6 aligns with the transformed course in hindsight.
%One side effect of updating the lectures that supplemented the lab activities was an increased reliance on published research papers from professional physics journals.
%This was done merely as a pedagogical ``best practice.''
%Nonetheless, it seems to have contributed to a favorable positive shift in students' attitudes and beliefs about scientific journal articles.
%
%Item 14, about agency in thinking up questions to investigate, was not a focus of the transformation.
%Students had no greater agency in choosing the research topics they would investigate during their lab activities, in fact if anything, students in the transformed course had less choice of lab activity than those in the original course.
%Further research is needed to understand why the data from the transformed course shows a significant favorable shift in item 14.
%
%Lastly, items 24 and 30 showed significant unfavorable shifts in the transformed course compared to the original.
%Neither of those items were a focus of the course transformation, either explicitly or otherwise.
%Again, further research is needed to understand these shifts in the transformed course.

\mysection{Conclusions and Outlook}
Compared to the original course, our transformed course did not yield a statistically different distribution of overall E-CLASS post-test scores.
However, on particular items related to the goals of the transformation, three showed statistically significant post-test differences after the transformation.
These items related to views about confirming know results, about measurement uncertainty, and about working in groups.
The three shifts were towards more expert-like responses.

%These findings provide an answer the research question posed above.
%While the course transformation discussed in this work does not affect the overall E-CLASS score compared to students who had taken the original course, we observe statistically significant and mostly favorable shifts in average scores on seven particular E-CLASS items when comparing data from before and after the transformation.
%Four of the items with favorable shifts align with particular aspects of the course which were changed in the transformation, and three of them correspond with explicit learning goals or design principles that were identified early in the transformation process.
%
%The remaining three items with statistically significant shifts, one favorable and two unfavorable, require further study to understand.
Future work will involve a deeper analysis of E-CLASS data from both the original and the transformed course, and will compare E-CLASS findings to results from a variety of other assessment tools and research methods.
The transformed course will be further improved based on those findings, and provide opportunities for further study.
We expect that these data sets, when analyzed together, will provide a more complete picture of the effects of our transformation on student learning, and provide a blueprint for transforming and evaluating lab courses more generally.
%Further course improvements will also be informed by the experiences of course instructors, lab managers, and teaching assistants. 
%Future iterations of the course will be studied similarly to Spring 2018, including using E-CLASS.
%Those data sets might also provide more insight into the effects of the transformation on students' attitudes and beliefs about experimental physics.
%In conclusion, our course transformation had an overall favorable effect on particular aspects of student learning in the introductory lab course at CU.
%Our results contribute to a growing body of research identifying, measuring, documenting, and elevating the unique and essential learning that occurs in lab courses.
%This study is part of a larger effort to comprehensively understand the impacts of the transformed course using multiple instruments and research methods (cf. Refs. [8,9,12]). Beyond making statements about the particular course at CU, these studies will provide a blueprint for transforming and evaluating lab courses more generally.
As researchers further understand the nature of student experiences in physics lab courses, we will be able to build on the already valuable opportunities offered to students in these courses to further support the unique learning that is possible in physics labs.

%Future work
%~Continue to analyze other data (vague)
%~Continue to improve course

%Answered research question
%~And therefore labs are valuable!
\vspace{-1 em}
\acknowledgments{
\vspace{-1 em}
We acknowledge D. Bolton, R. Hobbs, C.West, S. Woody, M. Schefferstein, A. Ellzey, M. Dubson, and J. Bossert for their work in developing the transformed course. We also acknowledge D. Dounas-Frazer and L. R\'ios for their input during the course transformation, B. Wilcox for her help with E-CLASS data analysis, and the many student testers for their contributions to improving the labs. This work is supported by the NSF under grant PHYS-1734006, the office of the Associate Dean for Education of the College of Engineering and Applied Science and the College of Arts and Sciences at the University of Colorado Boulder.}

\bibliography{references_ArXiV}

\begin{thebibliography}{14}%
\makeatletter
\providecommand \@ifxundefined [1]{%
 \@ifx{#1\undefined}
}%
\providecommand \@ifnum [1]{%
 \ifnum #1\expandafter \@firstoftwo
 \else \expandafter \@secondoftwo
 \fi
}%
\providecommand \@ifx [1]{%
 \ifx #1\expandafter \@firstoftwo
 \else \expandafter \@secondoftwo
 \fi
}%
\providecommand \natexlab [1]{#1}%
\providecommand \enquote  [1]{``#1''}%
\providecommand \bibnamefont  [1]{#1}%
\providecommand \bibfnamefont [1]{#1}%
\providecommand \citenamefont [1]{#1}%
\providecommand \href@noop [0]{\@secondoftwo}%
\providecommand \href [0]{\begingroup \@sanitize@url \@href}%
\providecommand \@href[1]{\@@startlink{#1}\@@href}%
\providecommand \@@href[1]{\endgroup#1\@@endlink}%
\providecommand \@sanitize@url [0]{\catcode `\\12\catcode `\$12\catcode
  `\&12\catcode `\#12\catcode `\^12\catcode `\_12\catcode `\%12\relax}%
\providecommand \@@startlink[1]{}%
\providecommand \@@endlink[0]{}%
\providecommand \url  [0]{\begingroup\@sanitize@url \@url }%
\providecommand \@url [1]{\endgroup\@href {#1}{\urlprefix }}%
\providecommand \urlprefix  [0]{URL }%
\providecommand \Eprint [0]{\href }%
\providecommand \doibase [0]{http://dx.doi.org/}%
\providecommand \selectlanguage [0]{\@gobble}%
\providecommand \bibinfo  [0]{\@secondoftwo}%
\providecommand \bibfield  [0]{\@secondoftwo}%
\providecommand \translation [1]{[#1]}%
\providecommand \BibitemOpen [0]{}%
\providecommand \bibitemStop [0]{}%
\providecommand \bibitemNoStop [0]{.\EOS\space}%
\providecommand \EOS [0]{\spacefactor3000\relax}%
\providecommand \BibitemShut  [1]{\csname bibitem#1\endcsname}%
\let\auto@bib@innerbib\@empty
%</preamble>
\bibitem [{\citenamefont {Koponen}(2007)}]{Koponen2007}%
  \BibitemOpen
  \bibfield  {author} {\bibinfo {author} {\bibfnamefont {I.~T.}\ \bibnamefont
  {Koponen}},\ }\href {\doibase 10.1007/s11191-006-9000-7} {\bibfield
  {journal} {\bibinfo  {journal} {Science {\&} Education}\ }\textbf {\bibinfo
  {volume} {16}},\ \bibinfo {pages} {751} (\bibinfo {year} {2007})}\BibitemShut
  {NoStop}%
\bibitem [{\citenamefont {Caballero}\ \emph {et~al.}(2018)\citenamefont
  {Caballero}, \citenamefont {Dounas-Frazer}, \citenamefont {Lewandowski},\
  and\ \citenamefont {Stetzer}}]{Caballero2018}%
  \BibitemOpen
  \bibfield  {author} {\bibinfo {author} {\bibfnamefont {M.~D.}\ \bibnamefont
  {Caballero}}, \bibinfo {author} {\bibfnamefont {D.~R.}\ \bibnamefont
  {Dounas-Frazer}}, \bibinfo {author} {\bibfnamefont {H.~J.}\ \bibnamefont
  {Lewandowski}}, \ and\ \bibinfo {author} {\bibfnamefont {M.~R.}\ \bibnamefont
  {Stetzer}},\ }\href
  {https://www.aps.org/publications/apsnews/201805/backpage.cfm} {\bibfield
  {journal} {\bibinfo  {journal} {APS News Back Page}\ }\textbf {\bibinfo
  {volume} {27}} (\bibinfo {year} {2018})}\BibitemShut {NoStop}%
\bibitem [{\citenamefont {{National Research
  Council}}(2012)}]{national2012discipline}%
  \BibitemOpen
  \bibfield  {author} {\bibinfo {author} {\bibnamefont {{National Research
  Council}}},\ }\href {\doibase 10.17226/13362} {\  (\bibinfo {year} {2012}),\
  10.17226/13362}\BibitemShut {NoStop}%
\bibitem [{\citenamefont {{AAPT Committee on
  Laboratories}}(2014)}]{kozminski2014aapt}%
  \BibitemOpen
  \bibfield  {author} {\bibinfo {author} {\bibnamefont {{AAPT Committee on
  Laboratories}}},\ }\href
  {https://www.aapt.org/Resources/upload/LabGuidlinesDocument_EBendorsed_nov10.pdf}
  {\bibfield  {journal} {\bibinfo  {journal} {Am Assoc Phys Teach}\ } (\bibinfo
  {year} {2014})}\BibitemShut {NoStop}%
\bibitem [{\citenamefont {Wilcox}\ and\ \citenamefont
  {Lewandowski}(2018)}]{Wilcox2018}%
  \BibitemOpen
  \bibfield  {author} {\bibinfo {author} {\bibfnamefont {B.~R.}\ \bibnamefont
  {Wilcox}}\ and\ \bibinfo {author} {\bibfnamefont {H.~J.}\ \bibnamefont
  {Lewandowski}},\ }\href {\doibase 10.1119/1.5009241} {\bibfield  {journal}
  {\bibinfo  {journal} {Am. J. Phys.}\ }\textbf {\bibinfo {volume} {86}},\
  \bibinfo {pages} {212} (\bibinfo {year} {2018})}\BibitemShut {NoStop}%
\bibitem [{\citenamefont {Dounas-Frazer}\ and\ \citenamefont
  {Lewandowski}(2017)}]{Dounas-Frazer2017}%
  \BibitemOpen
  \bibfield  {author} {\bibinfo {author} {\bibfnamefont {D.~R.}\ \bibnamefont
  {Dounas-Frazer}}\ and\ \bibinfo {author} {\bibfnamefont {H.~J.}\ \bibnamefont
  {Lewandowski}},\ }\href {\doibase 10.1103/PhysRevPhysEducRes.13.010102}
  {\bibfield  {journal} {\bibinfo  {journal} {Phys. Rev. PER}\ }\textbf
  {\bibinfo {volume} {13}},\ \bibinfo {pages} {010102} (\bibinfo {year}
  {2017})}\BibitemShut {NoStop}%
\bibitem [{\citenamefont {Dounas-Frazer}\ and\ \citenamefont
  {Lewandowski}(2016)}]{Dounas-Frazer2016b}%
  \BibitemOpen
  \bibfield  {author} {\bibinfo {author} {\bibfnamefont {D.~R.}\ \bibnamefont
  {Dounas-Frazer}}\ and\ \bibinfo {author} {\bibfnamefont {H.~J.}\ \bibnamefont
  {Lewandowski}},\ }\href {\doibase 10.1119/perc.2016.pr.020} {\bibfield
  {journal} {\bibinfo  {journal} {PERC Proceedings}\ ,\ \bibinfo {pages} {100}}
  (\bibinfo {year} {2016})}\BibitemShut {NoStop}%
\bibitem [{\citenamefont {Pollard}\ \emph {et~al.}(2017)\citenamefont
  {Pollard}, \citenamefont {Hobbs}, \citenamefont {Stanley}, \citenamefont
  {Dounas-Frazer},\ and\ \citenamefont {Lewandowski}}]{Pollard2017}%
  \BibitemOpen
  \bibfield  {author} {\bibinfo {author} {\bibfnamefont {B.}~\bibnamefont
  {Pollard}}, \bibinfo {author} {\bibfnamefont {R.}~\bibnamefont {Hobbs}},
  \bibinfo {author} {\bibfnamefont {J.~T.}\ \bibnamefont {Stanley}}, \bibinfo
  {author} {\bibfnamefont {D.~R.}\ \bibnamefont {Dounas-Frazer}}, \ and\
  \bibinfo {author} {\bibfnamefont {H.~J.}\ \bibnamefont {Lewandowski}},\
  }\href {\doibase 10.1119/perc.2017.pr.073} {\bibfield  {journal} {\bibinfo
  {journal} {PERC Proceedings}\ ,\ \bibinfo {pages} {3}} (\bibinfo {year}
  {2017})}\BibitemShut {NoStop}%
\bibitem [{\citenamefont {Lewandowski}\ \emph {et~al.}(2017)\citenamefont
  {Lewandowski}, \citenamefont {Hobbs}, \citenamefont {Stanley}, \citenamefont
  {Dounas-Frazer},\ and\ \citenamefont {Pollard}}]{Lewandowski2017}%
  \BibitemOpen
  \bibfield  {author} {\bibinfo {author} {\bibfnamefont {H.~J.}\ \bibnamefont
  {Lewandowski}}, \bibinfo {author} {\bibfnamefont {R.}~\bibnamefont {Hobbs}},
  \bibinfo {author} {\bibfnamefont {J.~T.}\ \bibnamefont {Stanley}}, \bibinfo
  {author} {\bibfnamefont {D.~R.}\ \bibnamefont {Dounas-Frazer}}, \ and\
  \bibinfo {author} {\bibfnamefont {B.}~\bibnamefont {Pollard}},\ }\href
  {\doibase 10.1119/perc.2017.pr.056} {\bibfield  {journal} {\bibinfo
  {journal} {PERC Proceedings}\ ,\ \bibinfo {pages} {244}} (\bibinfo {year}
  {2017})}\BibitemShut {NoStop}%
\bibitem [{\citenamefont {Wilcox}\ and\ \citenamefont
  {Lewandowski}(2017)}]{Wilcox2017}%
  \BibitemOpen
  \bibfield  {author} {\bibinfo {author} {\bibfnamefont {B.~R.}\ \bibnamefont
  {Wilcox}}\ and\ \bibinfo {author} {\bibfnamefont {H.~J.}\ \bibnamefont
  {Lewandowski}},\ }\href {\doibase 10.1103/PhysRevPhysEducRes.13.020110}
  {\bibfield  {journal} {\bibinfo  {journal} {Phys. Rev. PER}\ }\textbf
  {\bibinfo {volume} {13}},\ \bibinfo {pages} {020110} (\bibinfo {year}
  {2017})}\BibitemShut {NoStop}%
\bibitem [{\citenamefont {Hu}\ \emph {et~al.}(2017)\citenamefont {Hu},
  \citenamefont {Zwickl}, \citenamefont {Wilcox},\ and\ \citenamefont
  {Lewandowski}}]{Hu2017}%
  \BibitemOpen
  \bibfield  {author} {\bibinfo {author} {\bibfnamefont {D.}~\bibnamefont
  {Hu}}, \bibinfo {author} {\bibfnamefont {B.~M.}\ \bibnamefont {Zwickl}},
  \bibinfo {author} {\bibfnamefont {B.~R.}\ \bibnamefont {Wilcox}}, \ and\
  \bibinfo {author} {\bibfnamefont {H.~J.}\ \bibnamefont {Lewandowski}},\
  }\href {\doibase 10.1103/PhysRevPhysEducRes.13.020134} {\bibfield  {journal}
  {\bibinfo  {journal} {Phys. Rev. PER}\ }\textbf {\bibinfo {volume} {13}},\
  \bibinfo {pages} {020134} (\bibinfo {year} {2017})}\BibitemShut {NoStop}%
\bibitem [{\citenamefont {Zwickl}\ \emph {et~al.}(2014)\citenamefont {Zwickl},
  \citenamefont {Hirokawa}, \citenamefont {Finkelstein},\ and\ \citenamefont
  {Lewandowski}}]{Zwickl2014a}%
  \BibitemOpen
  \bibfield  {author} {\bibinfo {author} {\bibfnamefont {B.~M.}\ \bibnamefont
  {Zwickl}}, \bibinfo {author} {\bibfnamefont {T.}~\bibnamefont {Hirokawa}},
  \bibinfo {author} {\bibfnamefont {N.}~\bibnamefont {Finkelstein}}, \ and\
  \bibinfo {author} {\bibfnamefont {H.~J.}\ \bibnamefont {Lewandowski}},\
  }\href {\doibase 10.1103/PhysRevSTPER.10.010120} {\bibfield  {journal}
  {\bibinfo  {journal} {Phys. Rev. PER}\ }\textbf {\bibinfo {volume} {10}},\
  \bibinfo {pages} {010120} (\bibinfo {year} {2014})}\BibitemShut {NoStop}%
\bibitem [{\citenamefont {Lewandowksi}\ \emph {et~al.}(2018)\citenamefont
  {Lewandowksi}, \citenamefont {Bolton},\ and\ \citenamefont
  {Pollard}}]{Lewandowksi2018}%
  \BibitemOpen
  \bibfield  {author} {\bibinfo {author} {\bibfnamefont {H.~J.}\ \bibnamefont
  {Lewandowksi}}, \bibinfo {author} {\bibfnamefont {D.}~\bibnamefont {Bolton}},
  \ and\ \bibinfo {author} {\bibfnamefont {B.}~\bibnamefont {Pollard}},\
  }\href@noop {} {\bibfield  {journal} {\bibinfo  {journal} {PERC Proceedings}\
  ,\ \bibinfo {pages} {submitted}} (\bibinfo {year} {2018})}\BibitemShut
  {NoStop}%
\bibitem [{\citenamefont {Lomax}\ and\ \citenamefont
  {Has-Vaughn}(2012)}]{Lomax2012}%
  \BibitemOpen
  \bibfield  {author} {\bibinfo {author} {\bibfnamefont {R.~G.}\ \bibnamefont
  {Lomax}}\ and\ \bibinfo {author} {\bibfnamefont {D.~L.}\ \bibnamefont
  {Has-Vaughn}},\ }\href@noop {} {\emph {\bibinfo {title} {{An Introduction to
  Statistical Concepts}}}},\ \bibinfo {edition} {3rd}\ ed.\ (\bibinfo
  {publisher} {Routledge},\ \bibinfo {year} {2012})\BibitemShut {NoStop}%
\end{thebibliography}%

\end{document}